\documentclass[fleqn,twoside]{article}
\usepackage{espcrc2}

\usepackage{graphicx}
\usepackage[figuresright]{rotating}

\newcommand{\AmS}{{\protect\the\textfont2
  A\kern-.1667em\lower.5ex\hbox{M}\kern-.125emS}}

\title{Minimum Bias Legacy  of Search Results }

\author{G. D'Agostini\address[MCSD]{Universit\`a di Roma ``La Sapienza'' and 
Sezione INFN di Roma1, Roma, Italy \\ 
        P.le A. Moro 2, I-00185 Roma, Italy\\
{\tt http://www-zeus.roma1.infn.it/\,$\tilde{ }$\,agostini.}}%
        \thanks{Talk given at the Seventh Topical Seminar on 
        ``The legacy of LEP and SLC '', 
        Siena, Italy,  8-11 October 2001.}
}

\begin{document}

\begin{abstract} 
The end of LEP and SLC is a good moment to review the 
way to summarize search results in order to 
exploit at best, in future analyses and speculations, 
the pieces of information
coming from all experiments. Some known problems with the 
usual way of
reporting results in terms ``CL limits'' are shortly recalled, 
and a plea is formulated to publish just parametrized likelihoods,
possibly rescaled to the asymptotic insensitivity limit level. 
\vspace{1pc}
\end{abstract}

\maketitle

\section{INTRODUCTION AND RATIONALE}\label{sec:desiderata}
This contribution starts from an observation which I hope
is shared by many other particle phycisists, and scientists in general. 
Science, and particle physics in the specific case, should be 
considered a global 
activity, with individuals cooperating at different levels.
Certainly, those who have played relevant role in the research should
be acknowledged and, in special exceptional cases, rewarded. 
But the outcome of the research should be considered, finally, 
intellectual property of the community which has allowed the 
research. 
This point of view implies that a result should be presented 
in such a way that the pieces of information provided by an 
experiment should be possibly exploited at best by other scientists. 
This does not mean that I agree with populistic ``do it yourself analysis'',
of which I have heard, as 
only those who know in depth detector performance
and phenomenology can provide sensible results. Therefore, the
problem is only that of making the result public in such a way that 
the data summaries can be used later in the most efficient way.  

As far as searches are concerned, the experiment could report 
a spectacular effect which fits easily in the rest of the 
knowledge (``the net of beliefs'') and  all members of the physics community 
believe to a discovery. In other, not controversial cases, there 
could be no hint at all, as it usually happens in most searches.
But there could be the case in which there are some indications.
The scientists are then in the doubt of making the claim, thus risking
their reputation, or keeping quite, thus loosing the chance. 
``The experiment was inconclusive, and we had to use statistics'',
said one once. But statistical methods are ``inconclusive'' by 
definition, if one interprets them rigorously: even 100 observed events 
over 1 expected background does not  necesseraly implies
that they are due to signal, though we all 
tend to believe so\ldots. Therefore, it is important 
to distinguish between what we rationally believe from
what the empirical facts teach us. 

Besides personal preference for prudence or to risk, it is a matter of fact
that the three situations sketched above (discovery, no effect and 
dubious hints) are not sharply separated. It is, therefore, important
to understand what is the ``optimal presentation'', following some
desiderata which can be summarised saying that 
the information contained in the experimental data should be
presented in the most powerful and unbiased way.
\begin{itemize}
\item
The result should not depend on whether 
one believes that there is no effect (i.e. giving a limit),
or that something has been found (a claim with an associated e.g. 
 mass, cross-section, etc).
\item
The pieces of evidence coming from different experiments 
should be combined in the most efficient way:
\begin{itemize}
\item
If many independent data sets each provide a little evidence
in favor of the searched-for signal, the combination
of all data should enhance that hypothesis. 
\item
If the indications are incoherent, their combination should provide 
a stronger constraint against that hypothesis.
\end{itemize}
\end{itemize}
These desiderata are not satisfied by the usual way of reporting
results of searches with ``CL limits''. There are also other kinds
of problems with  ``CL limits'', some of which will
be illustrated in next section. In particular, there is 
a problem of interpretation, which is the main cause of
false alarms in the past and of the spread of misleading 
information to the general public (consequently throwing bad shadows to
the reputation of particle physicists). 

\section{PROBLEMS WITH LIMITS. EXAMPLES FROM THE CONTACT INTERACTION SEARCHES}
The first time I was confronted with the problems of stating limits
was in occasion of an overall analysis of results on electron 
compositeness.\cite{Moriond} Analysing all results available at that
time I found the situation not really satisfactory. 
In many cases the operative procedure 
used to evaluate limits was not described, and sometimes the numerical 
values of published limits seemed to disagree with differential 
cross-section to which they referred. I considered this problem 
particulary serious because the procedure which I understood 
had most consensus at that time had technical intrinsic problems 
in about 10\% of the results, as it will be explained in a while. 
But out of the many douzens of results
(many reactions $\times$ many coupling $\times$ many experiments), 
no technical anomaly was reported. 

Other bad feature was the difficulty (or impossibility) to 
combine consistently the limits, with the consequent attitude
to quote only the larger limit, that often was nothing but the largest
statistical fluctuation, and not due to the
experiment having the higher resolution power for that channel.

\subsection{Costraining $\Lambda$ to infinity}
In a contact interaction  analysis, no effect is obtained when 
the scale  $\Lambda$ is infinity. Therefore, one looks in the
data for a compatibility of the measured  $\Lambda$ with infinity, 
thus providing a lower limit. Needless to say, 
a MINUIT minimization around infinity is not trivial at all, 
not to speak about the interpretation of results. As a consequence, 
I found that, at that times, there was much of kitchen to get 
a number to quote as lower limit out of
the MINUIT printout. The worst case
was an experiment which had got an \underline{upper} limit 
of 1.3 TeV for a $\Lambda$ in a certain coupling
(this was the number I got reanalysing their data, due
to a overfluctuation), 
but published
exactly that number as \underline{lower} limit. 

This technical problem can be overcome working with the 
conjugate quantity $\epsilon=1/\Lambda^2$, which should 
come about zero in case of no effect. The choice of the 
second power of $\Lambda$ is due to the observation that 
the new terms in the cross-section come with that weigth. 
As a consequence, any additive ``noise'' make $\epsilon$ Gaussian 
distributed around the ``true'' value. 
Reporting the result on $\epsilon$ is certainly a good 
empirical practice, also because the results can be,
in most cases,  easily 
combined with the standard weighted average formula. 
Moreover, the standard deviation of $\epsilon$ is an intrinsic 
property of the experiment, a kind of ``resolution power''
depending on quality of the detector, luminosity, 
sensitivity to a particular reaction
and  level of background. And, in fact, my proposal\cite{Moriond} 
was to use
simply $\sigma_\epsilon$ as measure of the resolution power of 
an individual experiment or of a combination of experiments.

The problem remains if we insist to report a CL limit. 
Calling $\epsilon_\circ$ the best fit value, and 
$\sigma_\epsilon$ the standard deviation (the latter is
related to the 
curvature - ``width'' -  of the $\chi^2$ parabola)
the standard 95\% lower limit for   $\Lambda^\pm$
is given by
\begin{eqnarray}
\Lambda^\pm &=& \frac{1}{\sqrt{1.64\sigma_\epsilon\pm\epsilon_\circ}}\,.
\end{eqnarray}
This implies that, if $|\epsilon_\circ|$ is approximately equal to 
$1.64\sigma_\epsilon$, then either limit becomes very large. 
If  $|\epsilon_\circ| > 1.64\,\sigma_\epsilon$ (10\% of cases) there are
problems with the standard procedure, including the fact that 
for either sign of coupling there should be an evidence, yielding
an upper limit for $\Lambda$. As a matter of fact, unwanted 
results are tamed using ``prescriptions'', including  
``Bayesian prescriptions'' -- a contradiction in terms, in my view.

Besides the details of procedures, it is clear that this kind
of approach can produce large limits just as statistical 
fluctuations, limits which have nothing to
do with the effective resolution power of the experiment for a particular
channel. It is also a 
matter of fact that the difficulty to combine limits obtained
in such a way resulted in a general tendency to believe 
that a larger limit would make the lower ones obsolete, though
the latter might result from higher resolution experiments.

\section{MISINTERPRETATION OF ``C.L. RESULTS''} 
A second problem with standard ``C.L.'' is their
interpretation, as it resulted from a survey I made 
in 1998\,\cite{maxent98}. 
It came out, in fact, that most particle physicists 
believed that, e.g., a 95\% C.L. lower bound on the Higgs mass meant 
that ``the mass of the
Higgs, provided it existed, has 95\% chance to be above the limit,
and 5\% chance to be below''.
So, the problem is: do we understand 
each other? Do we communicate the correct information to
the general public of tax payer which financed our research? 

Nowadays it is quite understood by many people that such probabilistic 
statements are erroneous and misleading, but there are still
people which use such wrong statements based on frequentistic
C.L.'s to report the results to the general public.\cite{Read} 
So, for example, the 2000 hint of a 115 GeV Higgs
was reported by a spokesperson of the LEP experiments saying that 
{\it ``It is a 2.6 sigma effect. 
So there's still a 6 in 1000 chance that what we are seeing are 
background events, rather than the Higgs''}\cite{Tully}. 
So, basically the problem persists, since
also those who agree on what CL's {\it should not} mean, but still stick 
on the frequentistic approach, have difficulty in explaining what those
results {\it do} mean.
In my opinion, the very reason of this matter of fact  
is that, from a genuine inferential point of view -- which is what matters
in Science \cite{Clifford} 
-- frequentistic CL's have {\it no meaning},\footnote{Here is the most recent
example I know of the series of misleading press releases
in which frequentistic {\it numbers} 
(i.e. without a precise inferential meaning) are translated 
into probabilistic statements: 
{\it ``The experimenters reported a three-sigma discrepancy in sin2qW, 
which translates to a 99.75 percent probability that the 
neutrinos are not behaving like other particles.''}\cite{NuTeV} 
}  
and only in some case, under some
hidden hypotheses, they can be related to sensible probabilistic statements.
This is the reason why I insist in my position that 
``the solution to the problem 
of confidence limits begins with removing the expression itself''.\cite{clw}

\section{BAYESIAN WAY OUT}
In my opinion, the solution is not to propose a new prescription with 
the hope that it will be adopted by some influential
friends, but rather to change 
radically the attitude. 
This means we should review critically how 
our beliefs about physics quantities or laws of nature are 
modified by empirical observation 
and, once we have understood this scheme, we should stick to it, using logic
instead of tradition and/or authority. The most powerful tool to learn
from data is -- and this is not merely my opinion -- a \underline{theorem}, 
which states how our beliefs
must be updated by new pieces of information. This is the undiscussed role 
of Bayes' theorem, on which there is agreement also by those who disagree
on the use of beliefs in physics (but this is a different story, on which 
there is little to discuss, since  Science 
is nothing but a collection of beliefs based on
empirical facts and intellectual constructions\ldots).
 
Referring to a physics quantity of
unknown value $\mu$, the Bayes' theorem can be shortly expressed as
\begin{equation}
f(\mu\,|\,\mbox{data,\,I}) 
\propto f(\mbox{data}\,|\,\mu,\mbox{I})\times f_\circ(\mu\,|\,\mbox{I})\,,
\end{equation}
where the three ingredients of the Bayesian inference are, from left to right, 
final pdf, likelihood and prior.  Note that 
$f()$ stands, in this approach, for the the pdf expressing 
the relative beliefs. 

It is self-evident that the likelihood has the role of re-shaping 
(re-weighting) the beliefs. 
This is a fundamental ingredient of inference, but not the only one. 
Usually it is 
considered ``more objective'' than priors, because 
it is easier to agree on the response
of a detector than on purely speculative values of $\mu$. Nevertheless, 
$f(\mbox{data}\,|\,\mu,\mbox{I})$ is a probability too, and, as such, 
tells how much we believe 
that some data could be observed, for every hypothesis on $\mu$. 
 
At this point, the usual objection is that ``there are the priors''. 
In my opinion, 
there is no real problem if we understand on the logically crucial, 
often practically irrelevant  
role of priors. Without them, it would be impossible to make the 
``probabilistic inversion'' 
from $f(\mbox{data}\,|\,\mu,\mbox{I})$ to $f(\mu\,|\,\mbox{data,I})$, 
which is the essence of the
probabilistic inference. As far as its practical role, it depends on 
the different problems, and,
more specifically, on the shape of the likelihood. 
This is the reason why I like
to classify the inferential problems into 
{\it closed} and {\it open} likelihood. 
\subsection{Closed likelihood}
The easy case is when, for a given set of data, the likelihood function,
${\cal L}(\mu) = f(\mbox{data}\,|\,\mu,I)$ is {\it closed} in both sides,
i.e.  ${\cal L}(\mu) \rightarrow 0$ when $\mu$ tends to the extremes 
of its physical region of definition (usually $-\infty<\mu<+\infty$ or
 $0<\mu<+\infty$). The best understood example of this case is 
a Gaussian response of the apparatus. 
Under this condition and, in particular, when the width of ${\cal L}(\mu)$
is narrow with respect to any rational prior knowledge (the 
so Savage's ``precise
experiment'' situation), the conclusions do not depend {\it practically}
(in the sense how the result is perceived) from any reasonable model 
of the prior. Obviously, there could be someone who has a pure 
mathematical approach to the problem and propose a fancy 
mathematical expression for the prior. But I do not think this is a problem
for  physicists (and, anyhow, I am very curious to meet a defender 
of such fancy prior to check how ready he/she is to defend it
 with a suitable combination of bets\ldots). 

\subsection{Open likelihood}
The question becomes really problematic when the likelihood
is {\it open} in either side, i.e. ${\cal L}(\mu)$ 
goes to a constant at the edges of the physical region (if it is open in both sides the 
experiment has little relevance, unless it helps in enhancing 
a certain region selected by other experiment having closed likelihood).
For example, in the Higgs search of the kind performed
at LEP, an infinite mass produces a non zero pdf of observing what
we do observe. 
This ``bad'' feature is due to background: whatever we observe,
we are never {\it certain}  that it is not due to 
background alone. If this is the case, we are never able to  
normalize the final pdf, unless we force the normalization
with a prior (or other empirical evidence,
like radiative correction in the specific problem of the 
Higgs~\cite{conPeppe},
which forbids (or strongly inhibits) high masses. 
It follows that the sensitivity to the prior is usually so critical
(for an extensive, introductory  discussion of
problem and proposed solution, see Ref.~\cite{conPia})
that one should refrain from publishing, and spread  to the general public,
probabilistic limits, or limit which are usually  
mis-interpreted as such (this is what
 happens practically always with CL's limits).

In the case of open likelihood, only the likelihood itself should be reported. 
This information can be easily combined with that coming from other experiments,
and satisfies the desiderata of Section \ref{sec:desiderata}. Alternatively, one could report the 
log-likelihood or the rescaled {\cal R} function proposed in Refs.~\cite{conPia,conPeppe}. 
This functions can be parametrized in a suitable way, and perhaps stored 
in web sites in the case of searches having the likelihood depending of several quantities. 
An example of published parametrized log-likelihood can be found in 
Ref.~\cite{zeus_ci}. 

An alternative to probabilistic, or CL bounds, to quantify with a single number
(or with contour lines in the case of multidimensional analysis) is to 
report the {\it sensitivity bound}. This quantity should give, though roughly,
the edge after which the experiment looses sensitivity to
the search ~\cite{conPia,clw}.
 
\section{CONCLUSION: A PLEA FROM A EU TAX PAYER}
In conclusion, my plea to LEP and SLC colleagues, as well as to all
other physicists involved in searches, is to report likelihoods
for all searched channel, so that the effort of all community 
can be used at best for all future analyses. 

I would like to remember that publishing likelihoods was, actually,
a point about which 
there was unanimous agreement 
in the January 2000 workshop on confidence limits held at CERN,
as explicitly asked 
by Massimo Corradi and recorded by Louis Lyons~\cite{Massimo}. 
It was, indeed, the only generally agreed conclusion of the 
workshop.~\footnote{In my point of view there were 
also another objective, though implicit, conclusion of that workshop. 
The fact itself that many ``specialists'', all educated 
on the same books of statistics, meet 
to try to solve self-evident problems, absurdities and contradictions 
originated by the approach followed by those books, 
means that the basic concepts of those books should be 
critically reviewed, and that the gurus of that school of
thought, responsible of such confusion, should simply 
keep silent for the rest of their life.}

On the light of the well known properties of the likelihood 
and on the agreement during that meeting, attended by representative
by all major experiments, it is surprising to realize that 
reporting likelihoods has not become the standard yet. Some 
say that I am too na\"\i ve, and that this will never 
happen, essentially for two reasons. First, publishing 
a likelihood is much more committing than just giving 
a single ``95\% CL limit'' in the region where other experiments 
report similar limits. Second, there are people specialized in 
combination of results which do use likelihoods, but are afraid to loose
this privileged position if likelihoods are available to any 
student. I hope it is not so, and that it is only due to 
some inertia. Therefore I am still optimistic (or na\"\i ve).

\end{document}